\documentclass[pdflatex,sn-mathphys-num]{sn-jnl}

\usepackage{graphicx}%
\usepackage{multirow}%
\usepackage{amsmath,amssymb,amsfonts}%
\usepackage{amsthm}%
\usepackage{mathrsfs}%
\usepackage[title]{appendix}%
\usepackage{xcolor}%
\usepackage{textcomp}%
\usepackage{manyfoot}%
\usepackage{booktabs}%
\usepackage{algorithm}%
\usepackage{algorithmicx}%
\usepackage{algpseudocode}%
\usepackage{listings}%

\usepackage{setspace}
\newgeometry{left=3.25cm,bottom=3.5cm}

\begin{document}

\title[Article Title]{Identifying Key Genes in Cancer Networks Using Persistent Homology}


\author*[1,2]{\fnm{Rodrigo Henrique} \sur{Ramos}}

\author[1]{\fnm{Yago Augusto} \sur{Bardelotte}}

\author[1]{\fnm{Cynthia de Oliveira Lage} \sur{Ferreira}}

\author[1]{\fnm{Adenilso} \sur{Simao}}

\affil*[1]{\orgname{University of São Paulo}, \orgaddress{\street{Av. Trab. São Carlense, 400}, \city{São Carlos}, \postcode{13566-590}, \state{São Paulo}, \country{Brazil}}}

\affil[2]{\orgname{Federal Institute of São Paulo}, \orgaddress{\street{Estrada Municipal Paulo Eduardo de Almeida}, \city{São Carlos}, \postcode{13565-820}, \state{São Paulo}, \country{Brazil}}}


\abstract{Identifying driver genes is crucial for understanding oncogenesis and developing targeted cancer therapies. Driver discovery methods using protein or pathway networks rely on traditional network science measures, focusing on nodes, edges, or community metrics. These methods can overlook the high-dimensional interactions that cancer genes have within cancer networks. This study presents a novel method using Persistent Homology to analyze the role of driver genes in higher-order structures within Cancer Consensus Networks derived from main cellular pathways. We integrate mutation data from six cancer types and three biological functions: DNA Repair, Chromatin Organization, and Programmed Cell Death. We systematically evaluated the impact of gene removal on topological voids ($\beta_2$ structures) within the Cancer Consensus Networks. Our results reveal that only known driver genes and cancer-associated genes influence these structures, while passenger genes do not. Although centrality measures alone proved insufficient to fully characterize impact genes, combining higher-order topological analysis with traditional network metrics can improve the precision of distinguishing between drivers and passengers. This work shows that cancer genes play an important role in higher-order structures, going beyond pairwise measures,  and provides an approach to distinguish drivers and cancer-associated genes from passenger genes.
}




\maketitle

\section{Introduction}

Cancer research has advanced significantly with the advent of high-throughput genomic data and the development of public databases. The availability of extensive genomic data has facilitated the development of computational and statistical methods in various fields, including the identification of cancer genes~\cite{stratton2009cancer}. A major challenge in analysing mutation data lies in distinguishing between passenger and driver mutations. Passengers are the result of random genetic alterations or evolutionary processes and do not contribute to cancer development. In contrast, driver mutations are responsible for the onset and progression of the disease, making them targets for therapeutic intervention and personalised medicine~\cite{stratton2009cancer,ostroverkhova2023cancer}. In this work, in addition to drivers and passengers, we also use the term ``cancer-associated genes'' to refer to genes with publications associating them with cancer but are not present in driver databases. 

Protein-protein interaction networks (PPIN) and pathway networks are graph-based models representing protein interactions within cells. PPIN encompasses the entire interactome, while pathway networks represent specific biological functions, working as subsets of the interactome~\cite{Pathway_analysis}. Numerous computational approaches use the topology of PPIN and pathway networks to investigate cancer-related phenomena, such as mutual exclusivity, and to identify driver genes~\cite{dimitrakopoulos2017computational,cutigi2021computational,cutigi2020approaches,deng2019identifying}.

Traditional network science measures mainly address individual nodes, communities, or the whole network. Although powerful, traditional methods can overlook the topological and structural significance of gene interactions between the node and community level. Given the limitations of traditional methods, the Persistent Homology (PH), a tool from algebraic topology, offers a novel way to analyse complex networks by capturing multi-dimensional features~\cite{masoomy2021topological,PH_PPIN_Cancer_Therapy_2015}. This approach enables the identification of higher-order structures in cancer networks, providing a deeper understanding of the roles that specific genes play in the context of these structures. 

The objective of this study is to employ PH to identify genes that form higher-order structures within cancer networks derived from pathway networks and to explore their relationship with cancer. We constructed Cancer Consensus Networks (CCNs) using data from six types of cancer and three major biological functions: DNA Repair, Chromatin Organisation, and Programmed Cell Death. To evaluate the impact of each gene on topological voids ($\beta_2$ structures) within the CCNs, we systematically removed individual nodes and analysed the resulting changes. We then examine the role of these impactful genes in cancer.

Our findings reveal that every gene that affects $\beta_2$ structures is either a known driver or a cancer-associated gene, with the potential to be new drivers. The CCNs were constructed using mutated genes from various types of cancer. Given that most mutations are passengers~\cite{ostroverkhova2023cancer,kumar2020passenger}, we emphasise that removing passenger genes does not affect $\beta_2$ structures. Furthermore, we evaluated these impactful genes (known drivers or genes associated with cancer) using traditional network science measures, highlighting how centrality metrics alone are insufficient to fully characterise them. Not all known drivers or cancer-associated genes in the CCNs impact the formation of $\beta_2$ structures. However, no passenger gene has such an impact. Our method exhibits high precision with low to medium recall in distinguishing between drivers, cancer-associated genes, and passengers. Integrating higher-order topological features with traditional measures makes it possible to achieve a more comprehensive understanding of a gene's role in cancer, which can be applied to evaluate candidate driver genes. 

This work is organised as follows. The next two sections, ``Cancer Mutation Data and Reactome's Super Pathways'' and ``Persistence Homology'', present the theoretical background for developing this research. The ``Methods'' section details the data pipeline and our use of PH to characterise genes in CCNs. The ``Results and Discussion'' section explores the removal of genes from networks, its impact on higher-order structures, and how drivers and cancer-associated genes play a critical role in it. Finally, we end our paper with the concluding remarks. An appendix with formal PH definitions is also included.

\section{Cancer Mutation Data and Reactome's Super Pathways }
Advancements in DNA sequencing technologies have led to the generation of extensive genomic data. In the field of cancer research, databases such as the International Cancer Genome Consortium (ICGC) and the Cancer Genome Atlas (TCGA) offer datasets containing gene and mutation data for various types of cancer. Among the available datasets, the Mutation Annotation Format (MAF) is a commonly used tab-delimited file that connects patient samples, genes, and mutations. Each patient has one or more samples, each sample containing multiple genes linked to one or more mutations. The MAF file is frequently utilised in exploratory and computational approaches to identify driver genes and study patterns of mutual exclusivity \cite{maftools_2016, deng2019identifying}. In this work, we used cancer data from TCGA. Since TCGA deidentifies and anonymises all patient information, ethical approval was not required for this research.

Mutated genes in MAF files can be classified as either drivers or passengers. Drivers are genes whose mutations are causally linked to cancer~\cite{stratton2009cancer}, with databases such as NCG~\cite{NCG} and IntOGen~\cite{intOGen} offering lists of well-established drivers. These databases update their lists as new evidence emerges regarding a gene's role in cancer. Passengers, on the other hand, are mutated genes present in the MAF file but are not relevant to cancer~\cite{stratton2009cancer}. Distinguishing between drivers and passengers remains a critical challenge in cancer genomics~\cite{ostroverkhova2023cancer}, leading to the development of numerous computational methods to identify new drivers~\cite{cutigi2020approaches}. In this paper, we consider the genes listed in these databases as ``known drivers'', with high confidence in their role in cancer. All other mutated genes can be passengers or cancer-associated genes with the potential to be new drivers.

Pathways consist of sets of genes that collaborate to produce specific biological functions. As pathways are subsets of the entire PPIN, they are considerably smaller and provide meaningful information on the biological roles of their genes~\cite{Pathway_analysis}. Recent research comparing human PPINs from various databases reveals substantial inconsistencies in their interactions and topological structures~\cite{ramos2024PPIN}. The same study shows that subnetworks, including pathway networks, are more consistent across different PPINs. These findings indicate that whole PPINs are incomplete and still evolving, with new interactions continuously being discovered, validated, or invalidated. In contrast, interactions within well-known pathways, such as those used in this study, are more established, making pathway networks a more reliable option compared to whole PPINs~\cite{ramos2024PPIN}.

The Reactome Knowledgebase (https://reactome.org) is an open access, peer-reviewed, expertly curated database focused on biological pathways~\cite{reactomeDataBase2022}. It offers a variety of online bioinformatics tools designed for the analysis and visualisation of pathway-related data. Additionally, Reactome includes a PPIN derived from its pathway networks~\cite{reactomePPIN2017}. In 2020, Reactome introduced ``Super Pathways'', a hierarchical organisation of pathways that begins with broad biological functions, such as Programmed Cell Death, and extends into more detailed subcategories, such as Apoptosis and Regulated Necrosis \cite{reactomeSuperPathways}. Reactome presents pathways as lists of genes, enabling the extraction of induced subgraphs from a PPIN to create Super Pathways Networks (SPNs), a procedure we explain in the Methods section.

\section{Persistence Homology}

Topological data analysis (TDA)~\cite{carlsson2009topology,chazal2017introduction,chazal2017high} is based on the principle that topology and geometry can be utilised to derive both qualitative and quantitative insights about the underlying structure of data. Topological methods rely on the definition of similarity or distance between data points, allowing comparisons between data sets that may exist in different coordinate systems.

Persistent Homology (PH)~\cite{zomorodian2005computing}, a method within TDA, examines the topological features of data on various scales. PH identifies and quantifies the size and number of structures, such as connected components, cycles, and voids, by constructing a corresponding topological space from the data. The PH framework is built upon some fundamental concepts: simplicial complexes, filtrations, chains, and boundaries. These concepts are formally defined and illustrated with examples in the appendix. In this section, we provide an overview of PH and demonstrate its application in network analysis.

\begin{figure}[!htp]
    \centering
    \includegraphics[width=0.8\textwidth]{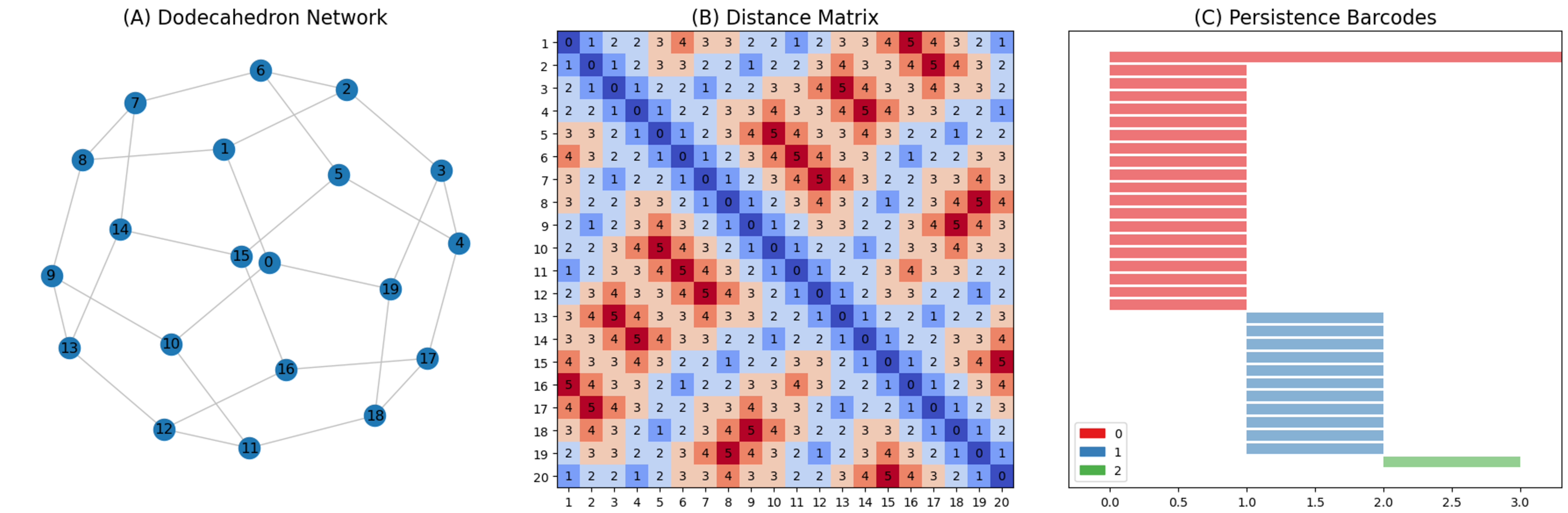}
    \caption{From network to persistence barcodes.}
    \label{fig:Distance Matrix}
\end{figure}

Typically PH is calculated over a point cloud, as exemplified in
Figure~\ref{fig:HP1} in the appendix. However, PH can also be computed over a network by defining a metric space based on a distance matrix calculated by pairwise distances between nodes. Figure~\ref{fig:Distance Matrix} demonstrates this process. Figure~\ref{fig:Distance Matrix}-(A) shows a network that resembles a dodecahedron, with 20 nodes and 30 edges. Figure~\ref{fig:Distance Matrix}-(B) shows a \(N \times N\) distance matrix calculated using the shortest path length between nodes. This matrix is the metric space used to calculate PH. Figure~\ref{fig:Distance Matrix}-(C) presents the Persistence Barcode, a plot normally used to visualise structures found during the PH. We will detail this in the next section.


\subsection{Persistence, Barcodes and Betti Numbers}
PH identifies the topological structures within the data. During the filtration step (explained in the appendix), structures are born at a given time and die at another. Significant structures persist longer than noise structures and are meaningful for characterising the data. Persistence barcodes represent the birth and death of topological structures across multiple scales. In Figure~\ref{fig:Distance Matrix}-(C), the bar colours represent different dimensions: red bars indicate connected components, blue bars indicate cycles (2-dimensional holes) and green bars indicate voids (3-dimensional holes). The X-axis of Figure~\ref{fig:Distance Matrix}-(C) shows the passage of time, i.e., the filtration process. Twenty red bars appear at time 0, and 19 persist until time 1, when the filtration process connects all loose connected components to one. This connection occurs at time 1 because the edges in Figure~\ref{fig:Distance Matrix}-(A) weight 1. At time 1, the dodecahedron faces are identified and persist for 1 tick of time. At time 2, a void is identified, representing the empty space inside the dodecahedron network. In summary, PH successfully identified the topological structures in Figure~\ref{fig:Distance Matrix}-(A), and the persistence barcode is a way to represent them. 

Betti numbers quantify the topological features of a space. Specifically, the $k$-th Betti number $\beta_k$ represents the number of $k$-dimensional holes in the data. $\beta_0$ counts the number of connected components, $\beta_1$ counts the number of cycles, and $\beta_2$ counts the number of voids. In Figure~\ref{fig:Distance Matrix}-(C), we have $\beta_0 = 20$, $\beta_1 = 11$, and $\beta_2 = 1$. As a polyhedron, the dodecahedron consists of 12 pentagonal faces. However, persistence homology identified only 11 cycles because not all faces contribute to distinct cycles. The edges of the ``missing'' cycle are shared with adjacent cycles, thereby not forming an independent cycle. Betti numbers provide a convenient method for quantifying the structures represented in Persistence Barcodes. In this work, we focus on using Betti numbers rather than barcodes, as our primary concern is the number of structures in the network and the impact individual genes have on them.

\subsection{Persistence Homology in Cancer Studies}
PH is an innovative tool in data science and has made contributions in many fields, such as network science, physics, chemistry, biology, and medicine~\cite{PH_OtherAplications_01,PH_OtherAplications_02,PH_OtherAplications_03,PH_OtherAplications_04,PH_OtherAplications_05,PH_OtherAplications_06}, thanks to its ability to analyse high-dimension datasets and extract meaningful features from complex data. 

In cancer studies, PH has been applied in various contexts, including image analysis, protein networks, gene expression networks, and point clouds. Specifically, PH has been used to evaluate prostate cancer in order to improve the Gleason grading system by capturing structure features independently of Gleason patterns. By computing topological representations of prostate cancer histopathology images, PH demonstrates the ability to group these images into unique groups through a ranked persistence vector. This method showed sensitivity to specific substructure groups within single Gleason patterns, offering a higher granularity than existing measures. The topological representations generated by PH could improve future approaches for better diagnosis and prognosis~\cite{PH_prostateCancer_2019}.

Furthermore, PH has been utilised in the study of protein interactions in the KEGG database to inform cancer therapy by analysing the correlation between Betti numbers and patient survival~\cite{PH_PPIN_Cancer_Therapy_2015}. In the context of gene expression networks, PH has been employed to examine gene interactions, uncovering structural features of the disease. It highlights significant deviations in the network topology between cancerous and healthy cells, emphasising the importance of cycles in cancer cells and voids in healthy cells~\cite{masoomy2021topological}.

Moreover, PH has been applied in tumour segmentation of Hematoxylin and Eosin stained histology images to enhance computer-aided diagnosis systems. This approach segments tumours in whole-slide images by analysing the degree of connectivity among nuclei through persistent homology profiles, outperforming convolutional neural networks~\cite{PH_TumorSegmentation_2016}. Lastly, PH has been used to characterise comparative genomic hybridisation profiles in breast cancer, providing a deeper understanding of chromosome amplifications and deletions in an individual's genome. The results were aligned with previous studies and distinguished between cancer recurrence frequencies in chemotherapy-treated and nontreated patient populations, highlighting the potential of PH in genomic data analysis~\cite{PH_BreastCancer_2010}.

\section{Methods}
We selected three SPNs, Chromatin Organisation (CHR), DNA Repair (DNA), and Programmed Cell Death (PCD), due to the roles these biological processes play in cancer development~\cite{Chromatin_organization_cancer,DNA_Repair_Cancer,Programmed_cell_death_cancer}. Furthermore, these networks exhibit a high proportion of known driver genes~\cite{ramos2021topological}, making them suitable for our study. Although other SPNs, such as Gene Expression and Signal Transduction, are also relevant to cancer, their extensive size, comprising over a thousand nodes, renders them computationally infeasible for analysis using the Vietoris-Rips complex in PH analysis due to the prohibitive combinatorial costs involved.

\begin{figure}[!htp]
    \centering
    \includegraphics[width=0.8\textwidth]{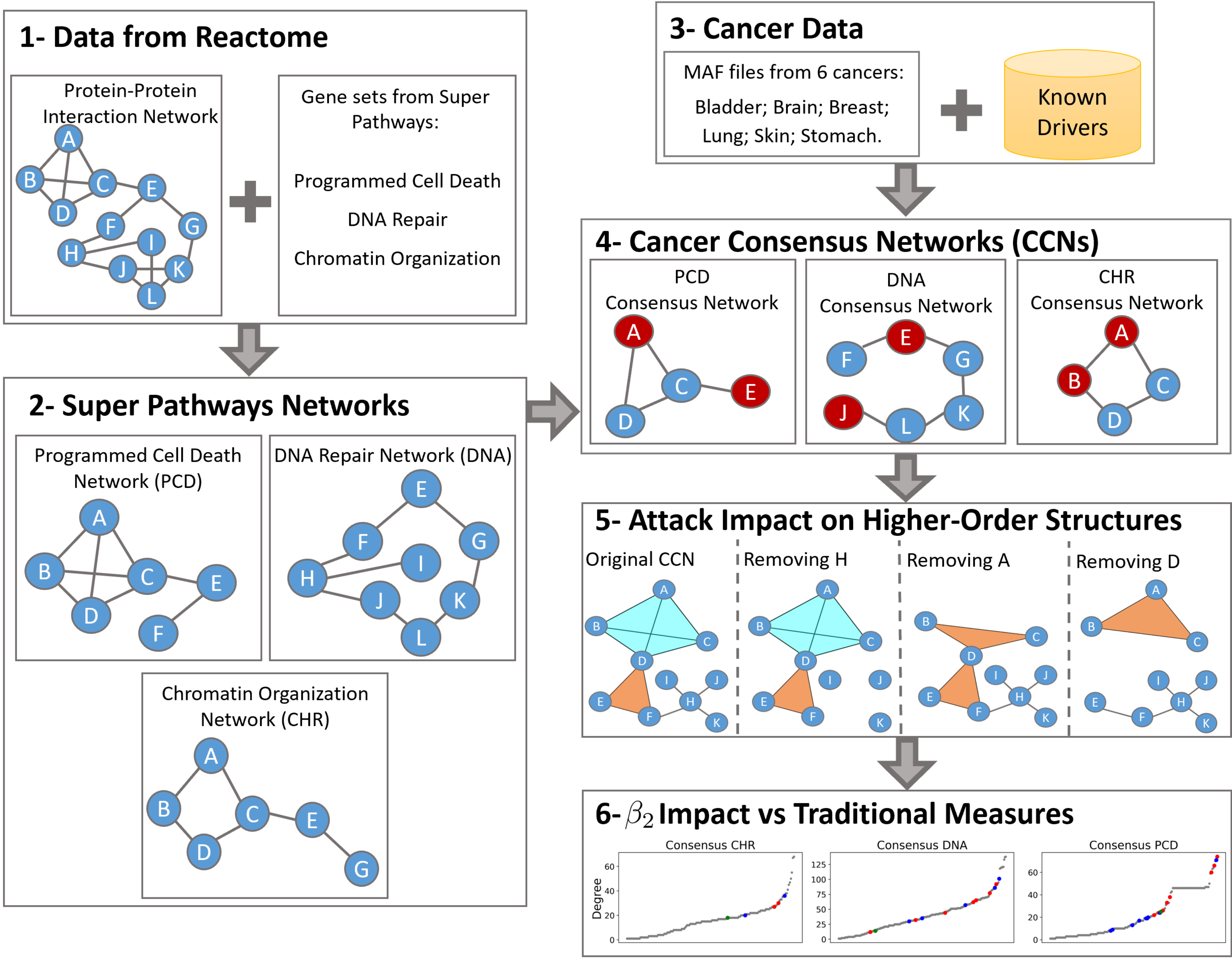}
    \caption{Data pipeline: We gather data from different databases to create Cancer Consensus Networks (CCNs), integrating data from three main biological functions with cancer-specific information. After that, we analyse the topological role of drivers and non-drivers in relation to their impact on higher-order structures.}
    \label{fig:Pipeline}
\end{figure}

The selected pathway networks represent the proteins and interactions present in normal and healthy cells. To associate these networks with cancer, we created the CCNs using mutation data from six types of cancer: Bladder, Breast, Head and Neck, Lung, Skin, and Stomach. Mutation data was obtained from MAF files in a TCGA pancancer study~\cite{MAFs_2018cell}. Figure~\ref{fig:Pipeline} shows the pipeline used in this work.

In the first step, we collected data from the Reactome PPIN and Reactome pathways. In the second step, we adopted a method similar to our previous research~\cite{ramos2021topological}, where we generated SPNs by extracting induced subgraphs from the Reactome PPIN using gene sets linked to Super Pathways. The third phase was conducted independently of the previous steps. We selected genes that were mutated in at least four of the six MAF files corresponding to different types of cancer. Furthermore, we identified known driver genes by considering the combined data from the intOGen~\cite{intOGen} and NCG~\cite{NCG} driver databases. Step four depends on steps two and three, since we use the selected mutated genes to extract induce subnetworks from each SPN, creating three CCNs. We also identify genes in the CCNs that are known drivers, represented in the plot as red nodes. The original SPNs for CHR, DNA and PCD contain 221, 300, and 206 nodes, respectively. Their corresponding CCNs reduced the nodes to 162 (73\%), 233 (78\%), and 170 (83\%). The number of driver genes in CHR, DNA, and PCD are 45, 46, and 26, respectively. In particular, the consensus networks retained at least 93\% of the original driver genes. Although the total number of nodes in the consensus networks decreased by approximately 22\% compared to the original SPNs, the reduction in driver genes was only 7\%. 

The fifth step in Figure~\ref{fig:Pipeline} summarises the analysis we performed to characterise nodes regarding their topological role in higher-order structures. It begins by calculating the PH for each CCN and recording the $\beta_2$ value using the Vietoris-Rips complex~\cite{chazal2017introduction}. In this work, we only focus on $\beta_2$ impact, since they are topologically more significant, are built using $\beta_1$, and their removal can increase the number of $\beta_1$.  In the example, the original CCN contains one cycle, formed by the nodes {D, E, F}, and one void, formed by the nodes {A, B, C, D}. Following this initial characterisation of the network, we systematically remove each node, one at a time, from the network and measure its impact on the $\beta_2$ value compared to the original CCN.

In the figure's example, removing node \textit{H} creates three new connected components, but does not affect any higher-order structures. \textit{H}'s impact can not be measured using PH, but can be measured by traditional network science measures, as previously done in the context of SPN and drivers~\cite{ramos2021topological}. On the other hand, removing node \textit{A} barely affects the network by traditional measures, but it has a relevant impact on higher-order structures. Node \textit{A} removal destroys a void ($\beta_2$) and creates a new cycle ($\beta_1$). Contrary to nodes \textit{H} and \textit{A}, node \textit{D} significantly impacts both traditional measures and higher-order structures.

The sixth and final step in Figure~\ref{fig:Pipeline} illustrates the second analysis performed to characterise the nodes. For each CCN, we calculate four centrality measures: degree, clustering, betweenness, and closeness. We then identify the position of the nodes that affected $\beta_2$ in the initial analysis. This step aims to compare the novel approach introduced in this paper, i.e., the impact of node on $\beta_2$, with traditional centrality measures.

\section{Result and Discussion}
The main objective of this work is to use PH to identify genes that form higher-order structures in CCNs and explore their relationship to cancer. By applying our proposed methodology, we assess the impact of each gene on the CCN's $\beta_2$ by individually removing nodes. Our results demonstrate that every node impacting $\beta_2$ structures is either a known driver or a gene associated with cancer, which potentially represents new drivers. The CCNs are constructed using mutated genes from various types of cancer. Given that most mutations are passengers~\cite{ostroverkhova2023cancer,kumar2020passenger}, we emphasise that removing passengers does not affect $\beta_2$ structures. In addition, we analyse these impactful genes (known drivers or cancer-associated genes) using traditional network science measures and discuss how centrality measures alone fail to fully capture them.

\subsection{Impact on $\beta_2$ by single node removal}
We calculated the PH for each CCN, identifying two $\beta_2$ structures in the CHR CCN, four $\beta_2$ structures in the DNA CCN, and ten $\beta_2$ structures in the PCD CCN. The PCD CCN, despite being the smallest network, exhibited the highest complexity in higher-order structures. Table~\ref{tab:singleAttackImpact} lists every gene that impacts $\beta_2$ structures in each CCN, highlighting in bold known drivers.

\begin{table}[!htp]
\caption{Impact on $\beta_2$ structures by single node removal. Bold names are known drivers.}
\label{tab:singleAttackImpact}
\centering
\begin{tabular}{|c|c|l|}
\hline
\textbf{CCN} & \textbf{$\beta_2$  Impact} & \multicolumn{1}{c|}{\textbf{GENES}}                                                \\ \hline
CHR          & -1                  & ACTL6A, BRMS1, \textbf{RELA}, \textbf{SMARCE1},   WDR77                                              \\ \hline
DNA          & -1                  & \textbf{ATM}, \textbf{EP300}                                                                         \\ \hline
DNA          & -2                  & \begin{tabular}[c]{@{}l@{}}\textbf{ABL1}, ACTL6A, \textbf{ATR}, \textbf{FANCD2}, \textbf{HERC2}, KAT5, PCNA, POLN,\\ RAD51, \textbf{XPA}, XRCC6\end{tabular}  \\ \hline
PCD          & -1                  & \begin{tabular}[c]{@{}l@{}} \textbf{AKT1}, APAF1, BAD, BIRC2, CASP1, \textbf{CTNNB1},  MAPT, \textbf{RIPK1},\\ ROCK1, \textbf{STAT3}, \textbf{STUB1}, \textbf{TNFSF10}\end{tabular}  \\ \hline
PCD          & -2                  & \textbf{HSP90AA1N}, \textbf{PTK2}                                                                    \\ \hline
PCD          & -3                  & \textbf{CASP3}, CASP6, \textbf{CASP8}                                                                \\ \hline
PCD          & -5                  & \textbf{TP53}  \\ \hline
\end{tabular}
\end{table}

CHR CCN is the least complex network, with five genes destroying one $\beta_2$ structure. In the DNA CCN, most impacting genes affected two $\beta_2$ structures. The PCD CCN, the most complex network, exhibited a different pattern, with the majority of impacting genes affecting only one $\beta_2$ structure. Five of the six genes that impacted more than one $\beta_2$ structure are known drivers. In particular, TP53, one of the most well-known genes in cancer research and frequently mutated across various types of cancer~\cite{guimaraes2002tp53}, stands out for its ability to independently destroy five $\beta_2$ structures. Most of the known drivers in the analysed CCNs did not impact $\beta_2$ structures. We hypothesise that these genes may be involved in even higher-dimensional structures, beyond $\beta_2$. However, the exponential computational cost of performing Vietoris-Rips filtration restricts such an analysis. This limitation suggests an avenue for future research to develop a filtration method specific to cancer networks that could reduce computational costs and enable the exploration of these higher-dimensional structures.


Table~\ref{tab:singleAttackImpact} lists 35 unique genes, of which 20 are identified as known drivers according to the combined data from the NCG and IntOGen databases. Table~\ref{tab:impactingGenes}, details these 35 impacting genes anWe provide the most recent publications for genes not found in driver databases, and the most recent publications associating them with cancer. In particular, all 15 genes not found in drivers database are drug targets or related to cancer. 

\begin{table}[!htp]
\caption{All 35 genes impacting $\beta_2$ structures in CCNs. 20 are known drivers listed in the NCG or IntOGen databases. The Literature column presents the most recent publication associating the remaining 15 genes with cancer.}
\label{tab:impactingGenes}
\begin{tabular}{|l|c|c|l|}
\hline
\textbf{Gene} & \textbf{NCG} & \textbf{IntOGen} & \textbf{Literature}                                                          \\ \hline
ABL1          & X            & X                &                                                                              \\ \hline
ACTL6A        & -            & -                & Association with Squamous Cell Carcinoma~\cite{ACTL6A_1}       \\ \hline
AKT1          & X            & X                &                                                                              \\ \hline
APAF1         & -            & -                & Melona drug target~\cite{APAF-1}                             \\ \hline
ATM           & X            & X                &                                                                              \\ \hline
ATR           & X            & X                &                                                                              \\ \hline
BAD           & -            & -                & Association with Triple-negative breast cancer~\cite{BADandCancer_2019} \\ \hline
BIRC2         & -            & -                & Head and Neck drug target~\cite{BIRC2}                      \\ \hline
BRMS1         & -            & -                & Metastasis suppressor in breast cancer~\cite{BRMS1}         \\ \hline
CASP1         & -            & -                & Association with Acute Myeloid Leukemia~\cite{CASP1}        \\ \hline
CASP3         & X            & -                &                                                                              \\ \hline
CASP6         & -            & -                & Association with Pancreatic cancer~\cite{CASP6}             \\ \hline
CASP8         & X            & X                &                                                                              \\ \hline
CTNNB1        & X            & X                &                                                                              \\ \hline
EP300         & X            & X                &                                                                              \\ \hline
FANCD2        & X            & X                &                                                                              \\ \hline
HERC2         & X            & -                &                                                                              \\ \hline
HSP90AA1      & -            & X                &                                                                              \\ \hline
KAT5          & -            & -                & Association with Hepatocellular carcinoma~\cite{KAT5}      \\ \hline
MAPT          & -            & -                & Association in Pan-Cancer~\cite{MAPT}                      \\ \hline
PCNA          & -            & -                & Drug target in multiple cancers~\cite{PCNA}                \\ \hline
POLN          & -            & -                & Association in Nasopharyngeal carcinoma~\cite{POLNandCancer_2022}        \\ \hline
PTK2          & X            & -                &                                                                              \\ \hline
RAD51         & -            & -                & Potential therapeutic target~\cite{RAD51}                   \\ \hline
RELA          & X            & X                &                                                                              \\ \hline
RIPK1         & X            & X                &                                                                              \\ \hline
ROCK1         & -            & -                & Association with Cancreatic cancer~\cite{ROCK1}             \\ \hline
SMARCE1       & X            & -                &                                                                              \\ \hline
STAT3         & X            & X                &                                                                              \\ \hline
STUB1         & X            & -                &                                                                              \\ \hline
TNFSF10       & X            & -                &                                                                              \\ \hline
TP53          & X            & X                &                                                                              \\ \hline
WDR77         & -            & -                & Association with Prostate cancer~\cite{WD77}               \\ \hline
XPA           & X            & -                &                                                                              \\ \hline
XRCC6         & -            & -                & Association with Lung Cancer Chemotherapy~\cite{XRCC6}      \\ \hline
\end{tabular}
\end{table}

The CCNs are extracted from SPNs using mutations from cancer patients, where the majority of mutations are passengers (i.e. not related to cancer). The results showed no $\beta_2$ impact upon removing passenger mutations, only consolidated known drivers or genes associated with cancer caused impact in higher-order structures.

\subsection{Impacting genes and centrality measures}

Taking into account traditional network science measures, drivers are known to have a high degree and work as hubs~\cite{driversHubs}, while some drivers genes have small degree~\cite{ramos2021topological}. Other works indicate that drivers can be categorised using additional centrality measures~\cite{driverCentrality1,driverCentrality2}. When characterising cancer driver genes, one of the key challenges lies in identifying drivers in the long tail of distributions associated with measures from protein networks and mutation data~\cite{cutigi2021computational}, as many methods are affected by ``ascertainment bias'', which tends to favour frequently mutated genes and network hubs~\cite{hierarchicalHotNet}. Here, we discuss whether genes impacting $\beta_2$ structures can be characterized using four centrality measures.

\begin{figure}[!htp]
    \centering
    \includegraphics[width=0.9\textwidth]{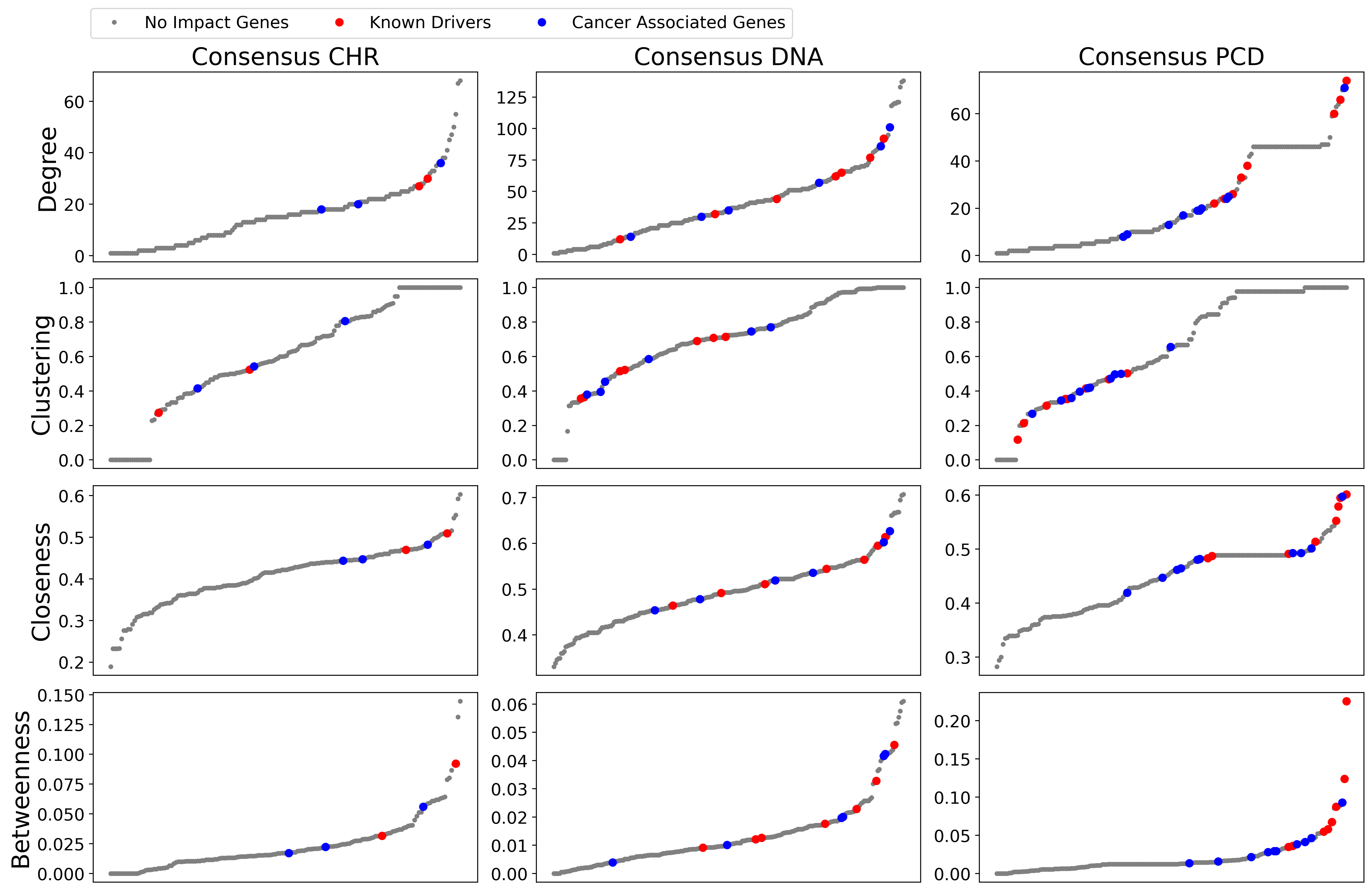}
    \caption{Centrality distributions for CCNs. Grey points represent genes whose removal does not affect $\beta_2$. Red and blue points indicate genes whose removal reduces $\beta_2$, with red points representing known drivers and blue points genes associated with cancer.}
    \label{fig:DegreeVSimpact}
\end{figure}

Figure~\ref{fig:DegreeVSimpact} displays the distributions of four centrality measures for all genes within each CCN. Grey points represent genes whose removal does not impact $\beta_2$, while red and blue points indicate genes whose removal decreases $\beta_2$, which correspond to the genes listed in Tables~\ref{tab:singleAttackImpact} and~\ref{tab:impactingGenes}. Red points are known drivers, and blue points are cancer-associated genes.

Overall, each centrality measure exhibits a similar distribution across the three CCNs, but the positions of the red and blue points vary. The CHR CCN has only five impacting genes, making it difficult to identify clear patterns. In this network, drivers and cancer-associated genes intermingle, occupying medium to high ranges in Degree, Closeness, and Betweenness. In the DNA CCN, with 13 impacting genes, the red and blue points are more evenly distributed in the middle, showing no clear distinction between drivers and cancer-associated genes, and they do not appear at the distribution extremes. Conversely, in the PCD CCN, drivers tend to occupy the top values in Degree, Closeness, and Betweenness, with low Clustering values. Additionally, there is a noticeable separation where known drivers tend to lead in these centrality measures, followed by cancer-associated genes.

Figure~\ref{fig:DegreeVSimpact} shows that no single centrality measure is sufficient to characterise the genes impacting $\beta_2$ structures. Although traditional centrality measures focus on nodes and edges within the network, they fail to capture the complexity of high-dimensional structures associated with these genes. This indicates that understanding the role of these genes requires going beyond basic centrality measures to account for the more complex, high-dimensional interactions and structures present in the network.

\section{Conclusion}

The study presents a novel approach to identifying known drivers and cancer-associated genes within cancer networks extracted from pathways using Persistent Homology. We constructed Cancer Consensus Networks by integrating mutation data from six types of cancer and three main biological functions. We measure the impact of removal of each gene in cancer networks with respect to its role in the construction of higher-order structures.  We complement the analysis using centrality measures to verify if traditional measures can capture the impacting genes.  The results demonstrate that only a few genes decrease the number of voids ($\beta_2$ structures). In particular, all impactful genes are established cancer drivers or cancer-associated genes, supported by existing literature, with the potential to be new drivers. Although not every driver or cancer-associated gene impacts $\beta_2$, no passenger gene does. The pipeline used in this work demonstrated high precision and low to average recall in distinguishing drivers from passengers. Although centrality measures alone do not fully characterise drivers and cancer-associated genes in CCNs, these genes generally exhibit low clustering and medium to high degree, closeness, and betweenness centrality values. This centrality profile, combined with the observation that no passenger mutations impact higher-order structures, can be utilized to evaluate candidate driver genes. Their topological characteristics can help determine their biological function as drivers or passengers.

\section*{Acknowledgements}

The authors acknowledge the financial support received from the Federal Institute of Sao Paulo (IFSP), the University of Sao Paulo (USP), the Sao Paulo Research Foundation (FAPESP), the Center for Mathematical Sciences Applied to Industry (CeMEAI), the Brazilian National Research and Technology Council (CNPq), and the Brazilian Federal Foundation for Support and Evaluation of Graduate Education (CAPES).

\section*{Declarations}

\subsubsection*{Ethical Approval}
The cancer data utilized in this study were sourced from TCGA. As TCGA de-identifies and anonymizes all patient information, ethical approval was not required for this research.


\subsubsection*{Availability of data and materials}
The mutation datasets are from a TCGA study~\cite{MAFs_2018cell} and can be download from cBioPortal.
\\
All code, input, and output files are on GitHub:\url{https://github.com/RodrigoHenriqueRamos/Identifying-Key-Genes-in-Cancer-Networks-Using-Persistent-Homology}


\bibliography{sn-bibliography}

\clearpage\section*{Appendix}
\subsection*{Simplicial Complex and Filtration}
\begin{figure}[!htp]
    \centering
    \includegraphics[width=0.8\textwidth]{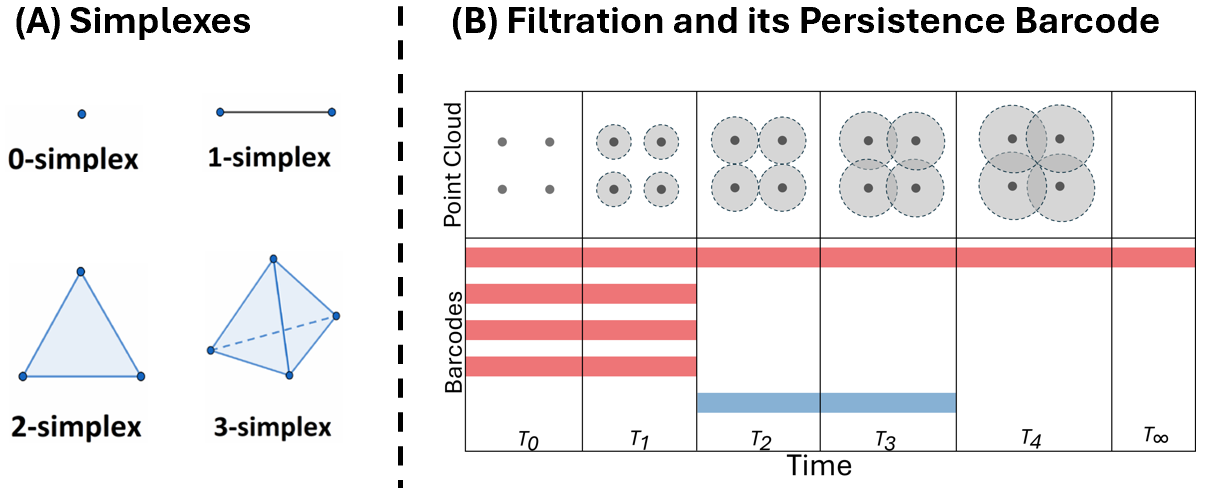}
    \caption{(A) Shows examples of simplexes of dimensions 0, 1, 2 and 3. (B) Presents the Vietoris-Rips filtration for a point cloud consisting of four equidistant points and the Persistent Barcode capturing the birth and death of topological structures.}
    \label{fig:HP1}
\end{figure}

A simplicial complex is a collection of simplices $\sigma=[0, ..., k]$ with dimension $k$. A 0-simplex corresponds to a vertex $[0]$, a 1-simplex to an edge $[0,1]$, a 2-simplex to a triangle $[0,1,2]$, a 3-simplex to a tetrahedron $[0,1,2,3]$, and so forth. Figure~\ref{fig:HP1}-(A) provides a visual representation of simplices. In a simplicial complex $K$, if a simplex $\sigma \subset K$, then every non-empty subset $\tau \subset \sigma$ is also a part of $K$. Additionally, two k-simplices in $K$ are either disjoint or intersect in a lower-dimensional simplex that is also contained within $K$.

To analyze the persistent homology of a dataset, a sequence of simplicial complexes $K_1 \subseteq K_2\subseteq ... \subseteq K_n$, known as a filtration, must be constructed. A common filtration is the Vietoris-Rips complex \cite{chazal2017introduction}, which constructs a topological space from a metric space using a distance $r$. By selecting a parameter $r>0$, we construct a simplicial complex $K_r$, where a k-simplex is included in $K_r$ if the distance between any two points forming the simplex is less than $2r$. It is important to note that if $r_2 \geq r_1$, then $K_{r1} \subseteq K_{r2}$. Figure~\ref{fig:HP1}-(B) exemplify a filtration over 4 points, and the birth and death of topological structures.

\subsection*{Chains and Boundaries}
To describe the topological properties in a simplicial complex $K_{r}$, we must utilize concepts from algebraic topology. In a simplicial complex $K$, for $k\geq0$, a k-chain $C_{k}$ is defined as a vector space whose basis consists of a set of k-simplices in $K$. The dimension of $C_{k}$ is determined by the number of elements in this basis. These vector spaces are composed of all linear combinations $c = \sum_{i} a_i \sigma_i$, where $a_i \in \mathbb{Z}p$ ($p$ being a prime integer) and the summation runs over all k-simplices $\sigma_i$ in $K$. Within this framework, the linear transformation $\partial_k: C_k\rightarrow C{k-1}$ can be introduced. This map, known as the boundary map, is defined as $\partial_k([0, ..., k])=\sum_{i=0}^{k} (-1)^i[0,...,\hat{i}, ..., k],$ where the notation $\hat{i}$ indicates the removal of that vertex. The boundary map applied to a simplex yields the alternating sum of the simplices along the boundary. Recall that the boundary of a k-simplex $\sigma$ is the union of the (k - 1)-subsimplices $\tau\subseteq \sigma$. Additionally, note that the composition $\partial_k$ $o$ $\partial_{k+1} = 0$. This implies that $im(\partial_{k+1})\subseteq ker(\partial_{k})$.

\begin{figure}[!htp]
    \centering
    \includegraphics[width=0.7\textwidth]{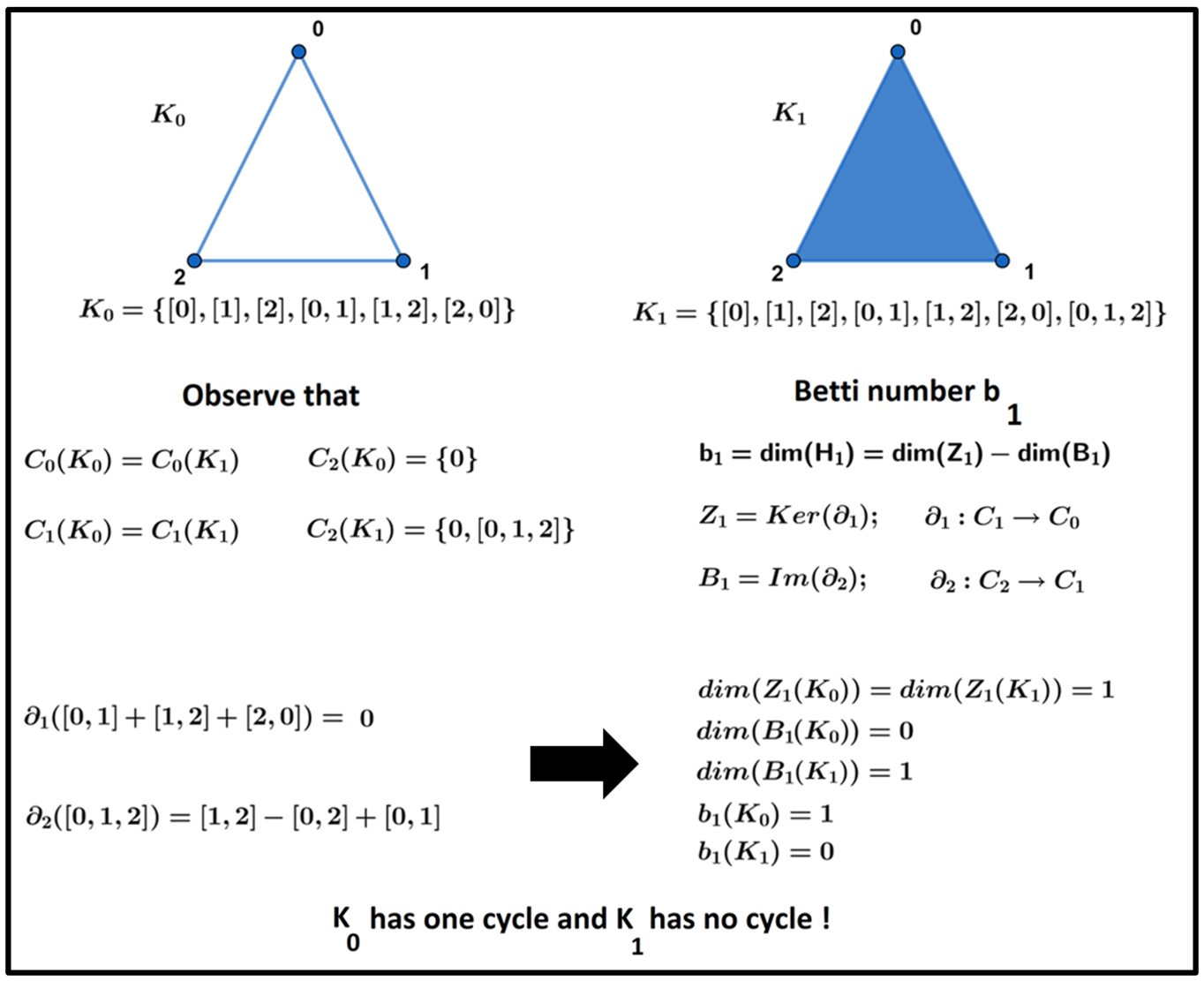}
    \caption{The Betti number $\beta_1$ for the complexes $K_0$ and $K_1$. Observe that the only element in $C_1$ that lies within the kernel of $\partial_1$ is $[0,1] + [1,2] + [2,0]$, leading to $dim(Z_1(K_0)) = dim(Z_1(K_1)) = 1$. Additionally, note that the boundary of the simplex $[0,1,2]$ forms the oriented triangle $[1,2]-[0,2]+[0,1]$. As a result, $dim(B_1(K_0)) = 0$, $dim(B_1(K_1)) = 1$, yielding $b_1(K_0) = 1$ and $b_1(K_1) = 0$. This indicates that $K_0$ contains a cycle, whereas $K_1$ does not.}
    \label{fig:Betti}
\end{figure}

Consider a chain complex $... \rightarrow C_{k+1} \xrightarrow{\partial_{k+1}} C_k \xrightarrow{\partial_{k}} C_{k-1} \rightarrow ... \rightarrow C_2 \xrightarrow{\partial_{2}} C_1 \xrightarrow{\partial_{1}} C_0 \xrightarrow{\partial_{0}} 0$, and define two subspaces of $C_k$ using the kernel and image of the boundary maps, namely $Z_k=ker(\partial_k)$ (k-cycles) and $B_k=im(\partial_{k+1})$ (k-boundaries). It is important to note that $B_k$ is a subspace of $Z_k$, and thus we can define $H_k = Z_k/B_k$, which represents the quotient of these vector spaces. The kth Betti number \cite{chazal2017introduction}, denoted by $\beta_k$, is defined as $\beta_k = dim(H_k) = dim(Z_k) - dim(B_k)$. Betti numbers quantify the number of ``holes'' in a simplicial complex $K$. Specifically, $\beta_0$ indicates the number of connected components in $K$, $\beta_1$ the number of cycles in $K$, $\beta_2$ the number of 2-dimensional holes, and more generally, $\beta_k$ represents the number of k-dimensional holes in $K$. Betti numbers are a topological invariant, meaning that topologically equivalent spaces share the same Betti numbers. Figure~\ref{fig:Betti} provides an example of calculating the Betti number $\beta_1$ for two distinct simplicial complexes.

\end{document}